\documentclass[twocolumn,prb,showpacs]{revtex4}

\usepackage{graphicx}

\usepackage{bm}

\begin{document}
\title{Entanglement production by independent quantum channels}

\author{\"Ors Legeza$^{1,2}$, Florian Gebhard$^{1}$, J\"org Rissler$^{1}$}
\affiliation{$^{1}$Fachbereich Physik, Philipps-Universit\"at Marburg,
D-35032 Marburg, Germany\\
$^{2}$Research Institute for Solid-State Physics and Optics,
H-1525 Budapest, Hungary}


\begin{abstract}
For the one-dimensional Hubbard model subject to periodic boundary conditions
we construct a unitary transformation between basis states so that
open boundary conditions apply for the transformed Hamiltonian. 
Despite the fact
that the one-particle and two-particle interaction matrices link nearest
and next-nearest neighbors only, the performance of the density-matrix
renormalization group method for the transformed Hamiltonian
does not improve. Some of the new interactions
act as independent quantum channels which generate the same level
of entanglement as periodic boundary conditions in the original formulation
of the Hubbard model. We provide a detailed analysis of these channels
and show that, apart from locality of the interactions, the performance of DMRG
is effected significantly 
by the number and the strength of the quantum 
channels which entangle the DMRG blocks.  
\end{abstract}

\maketitle

\section{Introduction}
\label{sec:intro}

The numerical density-matrix renormalization group (DMRG) method~\cite{White}
works best for lattice models with short-range interactions and
open boundary conditions. Non-localized versions have become
a major field of research, e.g., the DMRG in momentum 
space~\cite{Xiang,Gebhard-kdmrg,Legeza-ms} and in quantum 
chemistry~\cite{White-qc,Chan-qc,Mitrushenkov-qc,Legeza-dbss,Chan-qc1,Legeza-qc}.
For these applications, it is well established that the ordering of
`lattice sites' and the proper choice of basis states crucially
influence the convergence properties of the DMRG algorithm; for a review, 
see Refs.~[\onlinecite{Noack1,Dukelsky,Schollwoeck,Noack2}]. 
When Chan and Head-Gordon applied
a quantum-chemistry version of the DMRG (QC-DMRG)
to the calculation of the ground-state energy of selected 
molecules~\cite{Chan-qc},
they found that the DMRG leads to significantly
better results when lattice sites are re-ordered with the help of the
Cuthill--McKee algorithm~\cite{Cuthill}. However, using concepts
inherited from quantum information theory, 
it has been shown that the Cuthill--McKee algorithm fails to
generate an optimal ordering in general~\cite{Legeza-ms}.
In fact, it can lead to very bad configurations
which may even prevent the DMRG algorithm from converging to
the proper ground-state energy. 

The accuracy and convergence of the DMRG for given computer resources
is intimately related to the entanglement of the DMRG blocks during the
renormalization group step. Therefore, the von-Neumann entropy of the blocks
can be used to optimize the required computational 
resources~\cite{Legeza-ms,Legeza-qdc}.
The generation of block entropy as a function of system size was studied
in detail by various groups~\cite{Vidal,Korepin}. 
Recently, the entropy-approach was
extended in~[\onlinecite{Rissler1}] 
to include the two-site entropy profile. It suggests
a way to improve the criteria for the generation of
basis states and a proper ordering of the corresponding `lattice sites'.
The study of various orderings by brute-force algorithms confirmed
the best orderings as found from entropy-based methods but no
definite conclusions could be reached yet~\cite{Moritz-qc}. 

For lattice models, boundary conditions also have a strong influence
on the performance of the DMRG algorithm. 
When periodic instead of open boundary conditions
are used for the one-dimensional Hubbard model,
the block entropy 
increases significantly with system size~\cite{Legeza-qdc}.
In order to reduce numerical efforts to solve problems subject
to periodic boundary conditions, the matrix-product state description
has been introduced, for which, however, the  
interaction matrices become less sparse, and, 
thus, a true gain in performance  
could not be documented yet~\cite{Verstrate-pbc}.
Recent studies~\cite{Legeza-qdc,Legeza-deas}  
indicate that entanglement localization
and interaction localization actually compete and should be treated
on an equal footing. The central goal remains the development
of a standard procedure to find a basis state transformation 
for a given model which minimizes the block entanglement
and thereby optimizes the performance of the DMRG algorithm
in terms of required computational resources for a given demand on accuracy.

In this work we introduce a unitary basis transformation for
the one-dimensional Hubbard model with periodic boundary conditions
which results in a two-chain geometry with open boundary conditions
and couplings between nearest neighbors and next-nearest neighbors only.
Contrary to expectation, the performance of the DMRG algorithm 
does not improve. Our analysis shows that the transformation opens
new quantum channels which interfere with the kinetic-energy channel
and lead to a substantial entanglement between the DMRG blocks.

We organize our paper as follows. 
In section~\ref{sec:trafo} we describe briefly the Hubbard Hamiltonian
and the unitary transformation to the two-chain geometry with
localized interactions and open boundary conditions.  
In section~\ref{sec:numerics} we discuss 
our numerical procedure with an emphasis on the control of 
accuracy and the data analysis. We present our numerical DMRG results
in section~\ref{sec:results}.
We find that the DMRG procedure is more efficient for the Hubbard model
with periodic boundary conditions than for the transformed version
with open boundary conditions. We analyze this result in terms of 
the influence of independent quantum channels 
in section~\ref{sec:whywhywhy}. In particular, we show
that a super-site representation
for the two-chain geometry does not remedy the basic entanglement problem
of competing quantum channels.
We draw our conclusions in section~\ref{sec:conclusions}.

\section{Basis state transformation}
\label{sec:trafo}

\subsection{Hubbard model}

We consider the one-dimensional Hubbard model with uniform nearest-neighbor 
hopping on a finite chain of $L_{\rm s}$~lattice sites
subject to periodic boundary conditions,
\begin{eqnarray}
\hat{H} &=& -t \hat{H}_{\rm T}+U\hat{H}_{U}\;\nonumber \\
\hat{H}_{\rm T} &=& \sum_{j=0,\sigma}^{L_{\rm s}-1} 
\left ( \hat{c}^{+}_{j,\sigma} \hat{c}^{\phantom{+}}_{j+1,\sigma} 
+ \hat{c}^{+}_{j+1,\sigma} \hat{c}^{\phantom{+}}_{j,\sigma} \right) \; ,
\nonumber \\
\hat{H}_{U} &=& \sum_{j=0}^{L_{\rm s}-1} 
\hat{n}_{j,\uparrow} \hat{n}_{j,\downarrow}\;,
\label{eq:hub-pbc}
\end{eqnarray}
where $\hat{c}^{+}_{j,\sigma}$ ($\hat{c}^{\phantom{+}}_{j,\sigma}$) 
is the creation (annihilation) operator for electrons with 
spin~$\sigma=\uparrow,\downarrow$ at site~$j$,
$\hat{n}_{j,\sigma}= \hat{c}^{+}_{j,\sigma}\hat{c}^{\phantom{+}}_{j,\sigma}$,
and $\hat{n}_j=\hat{n}_{j,\uparrow}+\hat{n}_{j,\downarrow}$
is the occupation number at site~$j$. 
Due to periodic boundary conditions we set $\hat{c}_{L_{\rm s},\sigma}
\equiv \hat{c}_{0,\sigma}$.
 
\begin{figure}[htb]
\includegraphics[scale=0.7]{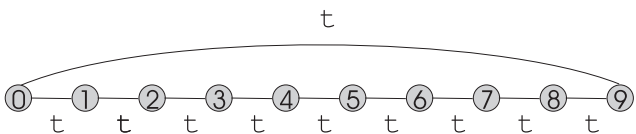}
\caption{Schematic plot of the Hubbard model with periodic 
boundary conditions for a chain with $L_{\rm s}=10$ lattice sites.
Solid lines denote the nearest-neighbor hopping 
while the on-site Coulomb interaction is shown by the gray shading.
Numbers indicate the lattice site indices.\label{fig:hub-pbc}}    
\end{figure}

The single-particle interaction matrix is given by 
$\hat{H}_{\rm T}$. We use the intersite hopping parameter~$t$ 
as unit of energy and set to $t=1$ in the following.
The two-particle interaction is given by $\hat{H}_{U}$ 
and $U$~is the strength of the on-site Coulomb interaction. 
The schematic plot of the model for a chain with $L_{\rm s}=10$
lattice sites with periodic boundary conditions 
is shown in Fig.~\ref{fig:hub-pbc}. 

It is evident from Fig.~\ref{fig:hub-pbc} and eq.~(\ref{eq:hub-pbc}) 
that the single-particle and the two-particle interaction matrices are 
diagonally dominated apart from the fact that 
$\hat{H}_{\rm T}$ has two off-diagonal terms due 
to the couplings between sites $j=0$ and $j=L_{\rm s}-1$
when periodic boundary conditions are employed.
These two terms leads to an enlarged bandwidth 
of $\hat{H}_{\rm T}$ and to a larger entanglement in the system 
as compared to the case of open boundary conditions.
Therefore, we should find a transformation which reduces the
bandwidth of $\hat{H}_{\rm T}$.
A reordering of lattice sites cannot lead to more localized interactions
as will be shown in section~\ref{sec:results}.
Thus, we need to apply an appropriate unitary transformation to new
basis states.

\subsection{Two-chain geometry}

Let us define the following unitary transformation for an even number of
lattice sites, 
\begin{eqnarray}
\hat{a}_{0,\sigma} \equiv \hat{c}_{0,\sigma} &,&
\hat{a}_{L_{\rm s}/2,\sigma} \equiv \hat{c}_{L_{\rm s}/2,\sigma}\;, \nonumber\\
\hat{a}_{j,\sigma} \equiv \sqrt{\frac{1}{2}} 
\left( \hat{c}_{j,\sigma} + \hat{c}_{L_{\rm s}-j,\sigma}\right)
&& \hbox{for } j=1,2,\ldots, \frac{L_{\rm s}}{2}-1\, ,    
\nonumber\\ 
\hat{b}_{j,\sigma} \equiv \sqrt{\frac{1}{2}} 
\left( \hat{c}_{j,\sigma} - \hat{c}_{L_{\rm s}-j,\sigma}\right)
&& \hbox{for } j=1,2,\ldots, \frac{L_{\rm s}}{2}-1\,. \nonumber \\ 
\label{eq:unitary}
\end{eqnarray}
The back-transformation reads for $j=1,2,\ldots,L_{\rm s}/2-1$ 
\begin{eqnarray}
\hat{c}_{j,\sigma} & \equiv & \sqrt{\frac{1}{2}} 
\left( \hat{a}_{j,\sigma} + \hat{b}_{j,\sigma}\right)\;,\nonumber \\
\hat{c}_{L_{\rm s}-j,\sigma} & \equiv & \sqrt{\frac{1}{2}} 
\left( \hat{a}_{j,\sigma} - \hat{b}_{j,\sigma}\right )\; .
\end{eqnarray}
The transformation is the result of a L\'anczos basis representation of
the kinetic energy which
starts from the state $|\Phi_0\rangle=\hat{c}_{0,\sigma}^+|{\rm vac}\rangle$.
In this way, all the operators~$\hat{a}_{j,\sigma}$
are generated. The operators~$\hat{b}_{j,\sigma}$ naturally follow
as the antisymmetric linear combinations of the
operators~$\hat{c}_{j,\sigma}$ and~$\hat{c}_{L_{\rm s}-j,\sigma}$.

\begin{figure}[htb]
\includegraphics[scale=1.0]{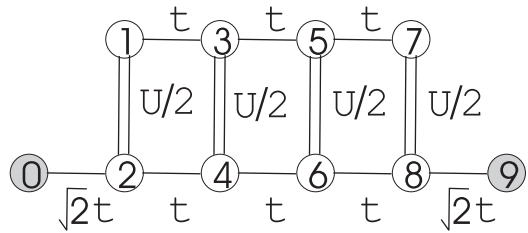}
\caption{Schematic plot of the transformed Hubbard model 
using the unitary transformation~(\ref{eq:unitary}).
Single solid lines denote single-particle couplings, double solid lines 
correspond to two-particle interactions. 
Shaded circles denote the on-site Coulomb interaction of strength~$U$
while empty circles correspond to strength $U/2$. 
Numbers indicate the lattice site indices.\label{fig:hub-transf}}    
\end{figure}

In terms of the new operators the kinetic energy becomes
\begin{eqnarray}
\hat{H}_{\rm T}  & = & \sum_{\sigma}
\sqrt{2}  \left[ 
\hat{a}_{0,\sigma}^{+} \hat{a}_{1,\sigma}^{\phantom{+}} 
+ 
\hat{a}_{L_{\rm s}/2,\sigma}^{+}\hat{a}_{L_{\rm s}/2-1,\sigma}^{\phantom{+}} 
+ {\rm h.c.} 
\right ] \nonumber \\ 
          && + \sum_{j=1,\sigma}^{L_{\rm s}/2-2} 
\left [ \hat{a}_{j,\sigma}^{+} \hat{a}_{j+1,\sigma}^{\phantom{+}} 
+ 
\hat{b}_{j,\sigma}^{+} \hat{b}_{j+1,\sigma}^{\phantom{+}} 
+ {\rm h.c.} 
\right ]\; .  \label{kintrans}
\end{eqnarray}
The geometry of the transformed model is shown sche\-mati\-cally 
in Fig.~\ref{fig:hub-transf}. As seen from the figure,
the transformed model displays
a two-chain geometry with open boundary conditions. 
Moreover, the kinetic energy only couples nearest-neighbor sites
of type~$a$ or~$b$. Thus, $\hat{H}_{\rm T}$ is diagonally dominated 
as for the case of the Hubbard model with open boundary conditions.

When we apply the unitary transformation 
to the two-particle interaction matrix we find
\begin{eqnarray}
\hat{H}_{U}  & = & 
n_{0,\uparrow}^a n_{0,\downarrow}^a 
+ n_{L_{\rm s}/2,\uparrow}^a n_{L_{\rm s}/2,\downarrow}^a 
\nonumber \\
& &+ \frac{1}{2} 
\left[
\hat{H}_{U, {\rm d}} + \hat{H}_{U, {\rm p}} +\hat{H}_{U, {\rm s}} 
\right]
\end{eqnarray}
with the local direct, pair-hopping, and spin-flip terms
\begin{eqnarray}
\hat{H}_{U, {\rm d}} &= & \sum_{j=1}^{L_{\rm s}/2-1}
\left( n_{j,\uparrow}^a + n_{j,\uparrow}^b \right)  
\left( n_{j,\downarrow}^a + n_{j,\downarrow}^b \right)
\; , \nonumber \\
\hat{H}_{U, {\rm p}} &=& \sum_{j=1}^{L_{\rm s}/2-1}  
\left( \hat{a}_{j,\uparrow}^{+} \hat{b}_{j,\uparrow}^{\phantom{+}}  
\hat{a}_{j,\downarrow}^{+} \hat{b}_{j,\downarrow}^{\phantom{+}}  
+ \hat{b}_{j,\uparrow}^{+} \hat{a}_{j,\uparrow}^{\phantom{+}}  
\hat{b}_{j,\downarrow}^{+} \hat{a}_{j,\downarrow}^{\phantom{+}}  
\right) \; , \nonumber \\ 
\hat{H}_{U, {\rm s}} &=& \sum_{j=1}^{L_{\rm s}/2-1}  
\left( \hat{a}_{j,\uparrow}^{+} \hat{b}_{j,\uparrow}^{\phantom{+}}  
\hat{b}_{j,\downarrow}^{+} \hat{a}_{j,\downarrow}^{\phantom{+}}  
+ \hat{b}_{j,\uparrow}^{+} \hat{a}_{j,\uparrow}^{\phantom{+}}  
\hat{a}_{j,\downarrow}^{+} \hat{b}_{j,\downarrow}^{\phantom{+}} 
\right)\; . 
\label{alltheUs}
\end{eqnarray}
The schematic plot of the two-particle interaction is also 
shown in Fig.~\ref{fig:hub-transf}. 
It is evident that we have transformed the Hubbard model 
with periodic boundary conditions into a two-band problem
with purely local interactions and nearest-neighbor electron transfers.

\subsection{One-dimensional representation}

\begin{figure}[htb]
\includegraphics[scale=0.6]{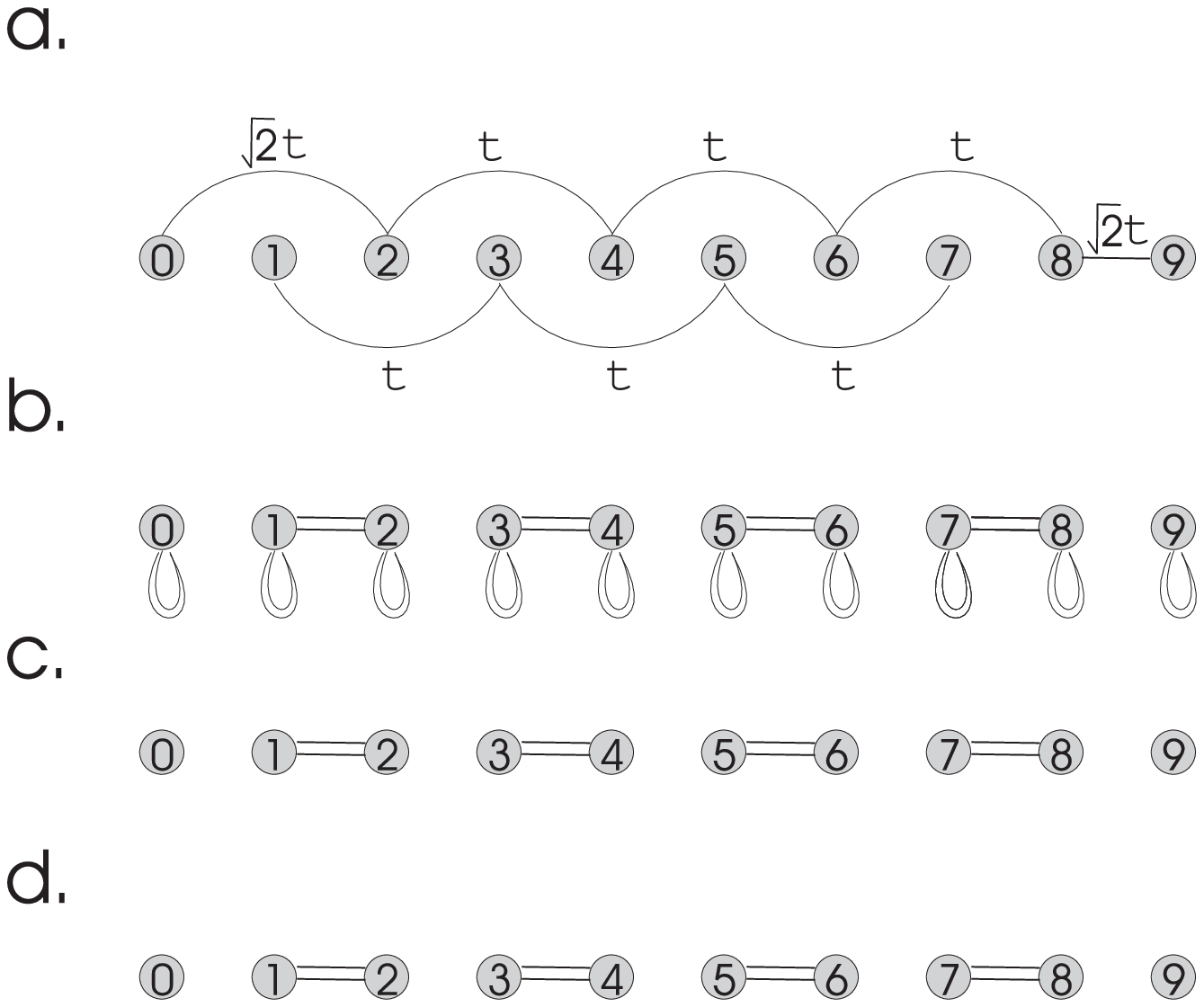}
\caption{One-dimensional representation of the single-particle electron 
transfers ${\hat H}_{\rm T}$ (a) and the two-particle interactions 
${\hat H}_{\rm U,d}$ (b), ${\hat H}_{\rm U,p}$ (c), ${\hat H}_{\rm U,s}$ (d) 
of the transformed Hubbard model. 
Loops in figure (b) denote on-site interactions.\label{fig:1d-ham}}
\end{figure}

The standard DMRG algorithm applies to one-band, i.e., single-chain geometries.
By ordering the sites of the two-chain geometry next to each other,
the Hamiltonian of the transformed model takes the form as
shown in Fig.~\ref{fig:1d-ham}.
The single-particle interaction matrix contains couplings between 
next-nearest neighbors. The two-particle interaction matrices 
contain on-site contributions via the direct term~$\hat{H}_{U, {\rm d}}$,
and nearest-neighbor interactions from $\hat{H}_{U, {\rm d}}$, 
the pair-hopping term~$\hat{H}_{U, {\rm p}}$, 
and the spin-flip term~$\hat{H}_{U, {\rm s}}$.
Nevertheless, all couplings remain local and the model is subject 
to open boundary conditions.
Therefore, we may expect that we can calculate ground-state properties
more efficiently in this formulation than we can for the 
Hubbard model~(\ref{eq:hub-pbc}) with 
periodic boundary conditions.

\section{Numerical procedure}
\label{sec:numerics}

\subsection{Controlling accuracy}

As in previous work~\cite{Legeza-dbss,Legeza-qdc}, we use
the dynamic block state selection (DBSS) procedure to control
the numerical accuracy of the DMRG method for different models.
When we apply the DMRG to the Hubbard model~(\ref{eq:hub-pbc})
with periodic boundary condition and the transformed model 
in the one-dimensional geometry of Fig.~\ref{fig:1d-ham},
we fix the quantum information loss~($\chi$) 
which is closely related to the relative error 
of the energy of the target state.
In this way, the block entropy as one of the
most relevant DMRG performance parameters 
can be monitored for different model Hamiltonians. 

We choose $\chi$ to make sure that the
maximum number of block states ($M_{\rm max}\simeq 3000$) 
that our program can handle is not reached during the calculations, i.e.,
we do not introduce an additional quantum information loss 
besides the truncation procedure based on $\chi$. 
We choose a small minimum number of block states $M_{\rm min}$ 
in order to make sure that its specific choice has negligible consequences
and yet ensures a reliable data analysis.
We use the entropy sum-rule as a criterion of convergence~\cite{Legeza-qdc}.
In general, five or six sweeps are carried out in order to make sure 
that the desired accuracy determined by $\chi$ has been reached. 

\subsection{Performance monitoring}
 
A natural quantity to measure the DMRG performance 
would be the CPU time. The CPU time, however, strongly depends on 
the CPU in use and a number of other technical issues. 
A software-related quantity to monitor is the block entropy
since it determines the number of blocks states required to reach the desired
accuracy for the given model and, thus, the speed of the DMRG calculations. 

\begin{figure}[t]
\includegraphics[scale=0.4]{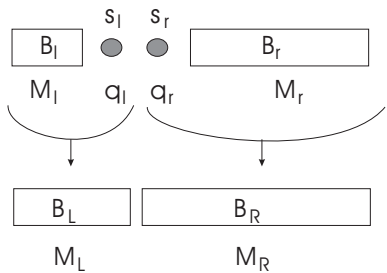}
\caption{Schematic plot of the system and environment block of DMRG. 
$B_{\rm l}$ and $B_{\rm r}$ denote the left and right blocks
of length $l$ and $r$, and of dimension $M_{\rm l}$ and $M_{\rm r}$, 
respectively, where $\bullet$ stands for the
intermediate sites ($s_{\rm l}$ and $s_{\rm r}$) 
with $q_{\rm l}$ and $q_{\rm r}$ degrees of freedom. 
The blocks $B_{\rm L} = B_{\rm l}\bullet$ and $B_{\rm R} = \bullet B_{\rm r}$ 
have dimensions $M_{\rm L}$ and $M_{\rm R}$, respectively.\label{fig:dmrg}}
\end{figure}

In this work we follow notations introduced in 
Refs.~[\onlinecite{Legeza-ms,Legeza-qdc,Rissler1}].
We decompose the total system into four subsystems. 
There are two sites, denoted by 
$s_{\rm l}$ and $s_{\rm r}$, with $q_{\rm l}$ and $q_{\rm r}$ 
degrees of freedom
between the left and right blocks, $B_{\rm l}$ and $B_{\rm r}$, of dimensions 
$M_{\rm l}$ and $M_{\rm r}$, respectively.
The blocks $B_{\rm L} = B_{\rm l}\bullet$, $B_{\rm R} = \bullet B_{\rm r}$ 
have dimension $M_{\rm L}$ and $M_{\rm R}$, respectively.
This configuration is shown in Fig.~\ref{fig:dmrg}.

The block entropies are denoted by 
$S_{\rm l}$, $S_{\rm L}$, $S_{\rm r}$, and $S_{\rm R}$. They, as well as 
the site entropies $S_{s_{\rm l}}$ and $S_{s_{\rm r}}$, are calculated
form the respective reduced subsystem density matrices~$\rho$ 
as $S=-{\rm Tr} \rho \ln \rho$. 
The number of degrees of freedom per site, $q_{\rm l}=q_{\rm r}\equiv q$, 
is $q=4$ for the Hubbard model and the transformed Hubbard model.

Apart from the entropies we monitor the Schmidt number $(\gamma)$
which counts the number of nonzero eigenvalues of the reduced
subsystem density matrix for each superblock partitioning,
\begin{equation}
|\Psi_T \rangle = \sum_{i=1}^{\gamma\leq\min(M_L,M_R)} 
\omega_i |\Psi^{(L)}_i\rangle \otimes |\Psi^{(R)}_i\rangle \; ,
\label{eq:Schmidtnumber}
\end{equation}
where $|\Psi_T\rangle$ is the wave function 
of the total system, $|\Psi^{(L)}_i\rangle$ and $|\Psi^{(R)}_i\rangle$
are bi-orthogonal basis states for the left and right blocks
with the condition $\sum_i \omega_i^2 = 1$.
The Schmidt number provides information  
about the entanglement of the 
subsystems when a pure target state is considered.
In our numerical analysis 
we determine $\gamma$ for a given quantum information loss $\chi$
and we demand $\omega_i>10^{-15}$ when we determine $\gamma$. 
Imposing this cutoff value 
induces some minor fluctuations in the Schmidt number
as a function of the strength of the quantum channels. 

\subsection{Total quantum information}

In order to compare more rigorously the various representations 
of a quantum system, we measure
the total quantum information, $I_{\rm tot}$,
encoded in the wave function.
To this end, we form all system blocks 
which contain $M_l=1$ to $M_l=L_{\rm s}$ lattice sites 
and sum up the quantum information gain of each 
renormalization group step~\cite{Legeza-qdc}. If no
truncation is applied, $I_{\rm tot}$ also equals 
the sum of the lattice-site entropies, 
i.e., $I_{\rm tot}=\sum_j S_{s_j}$. 
When we use the DBSS approach the error in $I_{\rm tot}$ is proportional 
to $L_{\rm s}\chi$.

\section{Results from DMRG}
\label{sec:results}

Let $E(L_{\rm s}, N_{\uparrow}, N_{\downarrow}, U)$ denote
the exact ground-state energy of the one-dimensional Hubbard model
for a finite chain with $L_{\rm s}$ lattice sites and $N_{\sigma}$ 
electrons with spin~$\sigma$ as a function of the interaction strength~$U$.
It can be obtained from the Bethe Ansatz~\cite{Bethe}.
In this work we study the paramagnetic half-filled case, 
$N_{\uparrow}=N_{\downarrow}=L_{\rm s}/2$ 
as a function of~$U$ for system sizes $L_{\rm s}\leq 64$.
All numerical data presented are from the results of 
the last DMRG sweep.

\subsection{Lattice site reordering}
\label{subsec:communication-length}

A reordering of lattice sites does not effect the total quantum correlation 
in the system. Therefore, when the model is solved exactly, $I_{\rm tot}$ is
a conserved quantity. The entanglement 
between the DMRG blocks, however, depends on
the number of quantum channels in between them and, thus,
the ordering of lattice sites has a major impact on the
performance of the DMRG method. 

\begin{figure}[htb]
\includegraphics[scale=0.6]{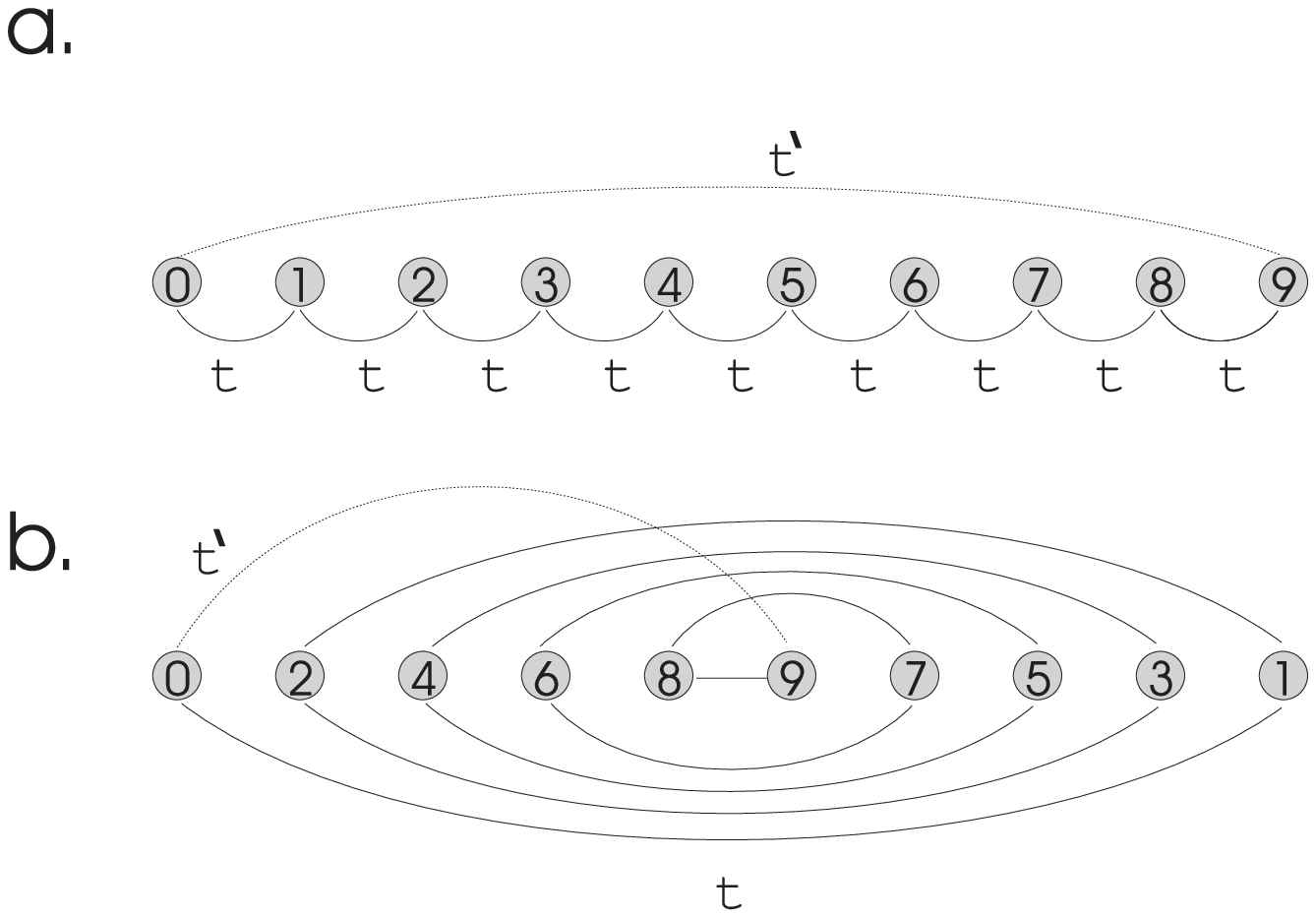}
\caption{Two extreme orderings for the Hubbard chain.\label{fig:hub-order}}
\end{figure}

For the Hubbard model two extreme 
configurations are shown in Fig.~\ref{fig:hub-order}.
In Fig.~\ref{fig:hub-order}~(a)
the ``communication'' between lattice sites $j=0$ and $j=L_{\rm s}-1=9$
is mediated by the neighboring
lattice sites in between them (strength~$t$) and 
by a direct channel (strength $t^{\prime}$).
The worst reordering is shown 
in Fig.~\ref{fig:hub-order}~(b) where the total length of
all communication paths between lattice sites $j=0$ and $j=L_{\rm s}-1=9$
is maximized. In order to quantify these statements we consider
the (total) ``communication length'',
\begin{equation}
C= \sum_{j=0}^{L_{\rm s}-1} |{\cal P}(j+1)-{\cal P}(j)| \; ,
\label{eqn:simplecommlength}
\end{equation}
where ${\cal P}$ permutes the $L_{\rm s}$ 
numbers $j=0,1,\ldots,L_{\rm s}-1$
into their new ordering. For example, in Fig.~\ref{fig:hub-order}~(b)
we set ${\cal P}_b(0)=0$, ${\cal P}_b(1)=L_{\rm s}-1$, 
${\cal P}_b(2)=2$, ${\cal P}_b(3)=L_{\rm s}-2$, and so on.
The standard ordering in Fig.~\ref{fig:hub-order}~(a) amounts to
$C^{\rm pbc}=2(L_{\rm s}-1)$ which can not be decreased by
any other ordering. For the reordering in Fig.~\ref{fig:hub-order}~(b),
the communication length is $C^{\rm max}=L_{\rm s}^2/2$.

For a more detailed analysis it is helpful to investigate the 
number of individual quantum channels between the left and right blocks
\begin{equation}
C_{\rm link}(l)= \begin{array}{l}
\hbox{[sum over all cross-links 
between $B_{\rm L}$ and }\\
\hbox{$B_{\rm R}$, where $(l+1)$ is the length of~$B_{\rm L}$.]}
\end{array}
\end{equation}
We have $C_{\rm link}^{\rm pbc}(l)=2$ 
in the configuration of Fig.~\ref{fig:hub-order}~(a) whereas 
$C_{\rm link}^{\rm max}(l)=L_{\rm s}-|2l+2-L_{\rm s}|$ in the configuration of
Fig.~\ref{fig:hub-order}~(b). From the number of quantum channels
we define the (total) communication length 
\begin{equation}
C=\sum_l C_{\rm link}(l)\;,
\label{eq:link-sum}
\end{equation}
which reduces to the expression~(\ref{eqn:simplecommlength}) for our example.

More generally, not only the number by also 
the strength and the type 
of the individual quantum channels between the left and right blocks play
an important role. In an obvious extension of~(\ref{eq:link-sum})
we may assign adjustable weight factors~$\gamma_{\rm channel}(l,U/t,\ldots)$
to each channel. Typically, the communication length~$C$
and the number of individual quantum channels 
at each link~$C_{\rm link}(l)$ are sufficient for a first assessment
of the entanglement of the system. The computational cost for one
DMRG iteration step is determined by $C_{\rm link}(l)$ since this number 
does not depend on the ordering of the lattice sites within the two blocks.
The overall cost of a full DMRG sweep, however, also depends on $C$ 
due to the relationship~(\ref{eq:link-sum}).
In future applications, $C$ may serve as a cost function 
to optimize the ordering (and the basis set).

\begin{figure}[htb]
\includegraphics[scale=0.37]{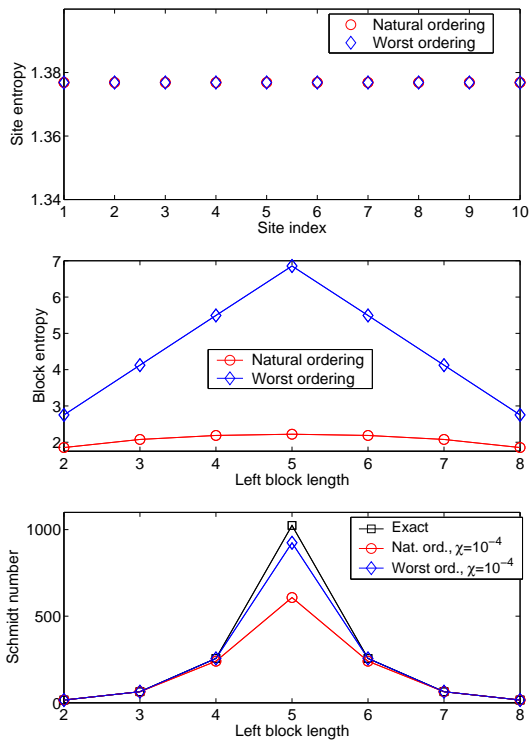}
\caption{Site entropy and block entropy  
for the half-filled Hubbard model for the two  
the ordering criteria shown in 
Fig.~\protect\ref{fig:hub-order} for $U=1$ and $\chi=0$ (exact calculation). 
The Schmidt number is also 
shown for $\chi=10^{-4}$ and $M_{\rm min}=4$. 
The lines are guides to the eyes.\label{fig:entropy}}
\end{figure}

In Fig.~\ref{fig:entropy} we show the site entropy, the block 
entropy, and the Schmidt number 
from exact DMRG calculations  
for $L_{\rm s}=10$ lattice sites at $U=1$
for the two configurations of Fig.~\ref{fig:hub-order}.
In order 
to make visible the differences in Schmidt numbers 
for small system sizes we include results for $\chi=0$ (exact calculation) and
$\chi=10^{-4}$ with $M_{\rm min}=4$.
After convergence all lattice sites posses the same entropy, 
$S_{s_i}\approx 1.377$,
and $I_{\rm tot}\approx 13.77$ is the same for both orderings. 
However, the block entropy 
is much larger for the configuration~\ref{fig:hub-order}~(b), i.e., 
one needs more computational resources to
solve the problem. Correspondingly, the  
Schmidt number $\gamma(\chi)$  for a given quantum information loss $\chi$
is larger for the worst ordering than for the natural ordering.
In fact, the block entropy and the Schmidt number closely follow
the number of cross-links between the left and right 
blocks, $C_{\rm link}(l)$, up to logarithmic corrections 
in the system size~\cite{Vidal,Korepin}.

There is no site ordering different from the 
configuration~\ref{fig:hub-order}~(a) which reduces the
number of cross-links below two, $C^{\rm min}_{\rm link}=2$,
and the communication length below $C^{\rm min}=2(L_{\rm s}-1)$. 
Thus, we conclude that the total quantum correlation 
in the system cannot be reduced by a reordering of the sites.
Only a basis-state transformation might offer a way to achieve
the desired entanglement reduction.

\subsection{Basis-state transformation}

\begin{figure}[t]
\includegraphics[scale=0.333]{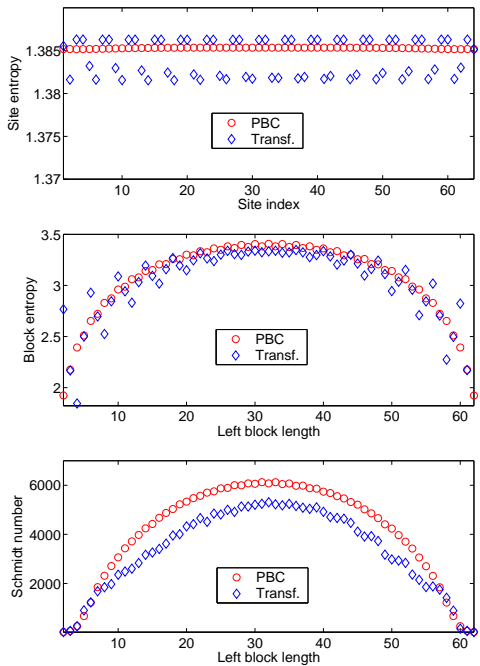}
\caption{Same as Fig.~\ref{fig:entropy} for 
the Hubbard model 
with periodic boundary conditions 
and for the transformed Hubbard model 
with open boundary conditions
for $\chi=10^{-4}$, $M_{\rm min}=64$ (PBC) and $M_{\rm min}=256$ (Transf.), 
$L_{\rm s}=64$, and $U=0$.\label{fig:hub-entropy-u0}}
\end{figure}

Since interactions are localized for the transformed Hubbard model 
we might expect that the entanglement in the system is reduced, 
and, thus, the problem can be solved more efficiently using DMRG. 
In Fig.~\ref{fig:hub-entropy-u0}  
we plot the site entropy, 
the block entropy, and the Schmidt number 
for the non-interacting Hubbard model and the
transformed model for $\chi=10^{-4}$, $M_{\rm min}=64$,
and $L_{\rm s}=64$ sites as a function of the number of DMRG sweeps. 
Note that the limit $U=0$ poses a non-trivial problem for 
the position-space DMRG.

For both models we determine the ground-state energy 
within the desired relative accuracy of better than $10^{-3}$.
The comparison of data points in Figs.~\ref{fig:hub-entropy-u0} 
shows, however, that the basis-state transformation~(\ref{eq:unitary})
did not lead to a significant improvement:
the site and block entropies as well as the Schmidt number 
are only marginally smaller than for the Hubbard model with periodic
boundary conditions. 

This result can again be understood from the
number of cross links and the
communication length in Fig.~\ref{fig:1d-ham}~(a).
The number of cross links is the same
for both representations, $C_{\rm link}^{\rm pbc}
=C_{\rm link}^{\rm transf}=2$, 
and the communication length is almost the same, too,
$C^{\rm transf}=2L_{\rm s}-5$ versus $C^{\rm pbc}=2L_{\rm s}-2$, 
see section~\ref{subsec:communication-length}.

\begin{figure}[htb]
\includegraphics[scale=0.333]{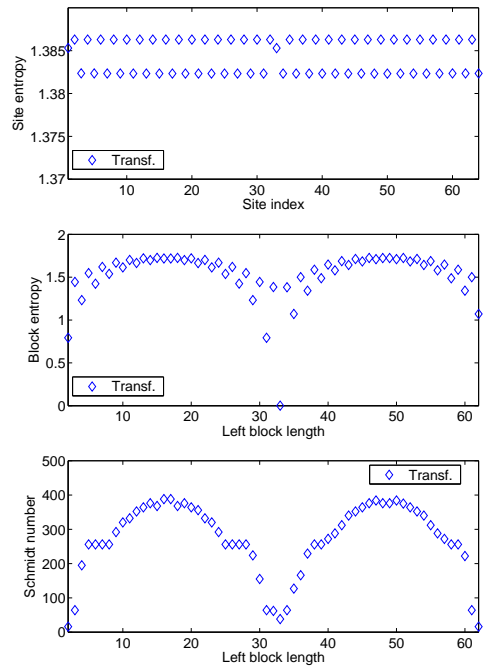}
\caption{Same as Fig.~\protect\ref{fig:hub-entropy-u0}
for the transformed Hubbard model 
with open boundary conditions for 
$U=0$, $\chi=10^{-4}$, and $M_{\rm min}=64$ in independent-chain 
geometry.\label{fig:hub-transf-entropy-u0-ord3}}
\end{figure}

In order to demonstrate the importance of the number of links 
and of the communication length,
we treat the two chains for $a$-electrons and $b$-electrons separately
as they decouple for $U=0$, see Fig.~\ref{fig:hub-transf}.
For this geometry, the DMRG result is shown 
in Fig.~\ref{fig:hub-transf-entropy-u0-ord3}.
The block entropy profile clearly shows the absence of
quantum correlations between the two independent chains.  
For this geometry we have 
$C_{\rm link}^{\rm two\hbox{-}chain}(l)=1$
for the $a$-chain and $b$-chain separately.
The communication length for both chains together
is $C^{\rm two\hbox{-}chain}=L_{\rm s}-2$. Therefore,
it is smaller by a factor of two  
than for periodic boundary conditions,
$C^{\rm pbc}=2 L_{\rm s}-2$. It is evident 
from Fig.~\ref{fig:hub-transf-entropy-u0-ord3} that the 
maximum of the block entropy has equally dropped by almost a factor of two,
from $3.3$ for periodic boundary conditions to $1.7$
for the two-chain geometry, and the Schmidt number 
has reduced by more than one order of magnitude.

\begin{figure}[htb]
\includegraphics[scale=0.333]{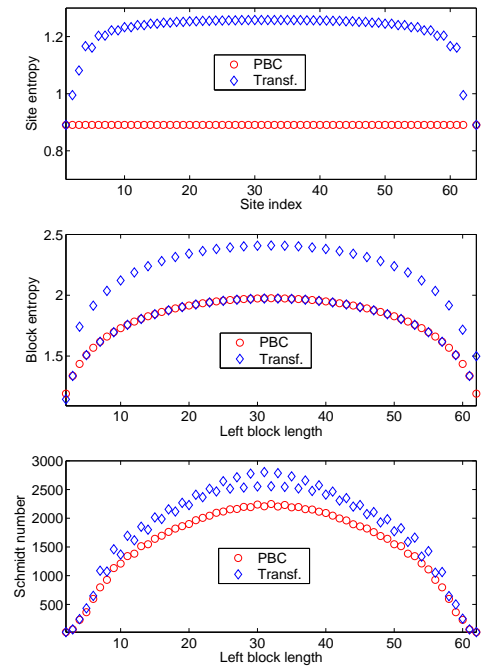}
\caption{Same as Fig.~\protect\ref{fig:hub-entropy-u0}
for $U=10$ and $\chi=10^{-5}$. \label{fig:hub-entropy-u10}}
\end{figure}

The situation drastically changes when the Hubbard interaction is 
switched on. 
As shown for $U=10$ in Fig.~\ref{fig:hub-entropy-u10},
the site entropy and the block entropy 
are actually {\sl larger\/} for the transformed Hubbard model 
with open boundary conditions so that {\sl more\/} block states 
are required to reach the same accuracy as in the original formulation 
of the Hubbard model with periodic boundary conditions. 
Note that the oscillation in the block entropy is related to the 
dimer configuration of the Coulomb interaction, i.e.,
the number of bonds between DMRG blocks $C_{\rm link}^{\rm transf}(l)$
oscillates between two and eight.
Therefore, the block entropy of the two representations is the same
for every second RG iteration step for which there are only the two
$t$-type channels between the DMRG blocks. 

\begin{figure}[tb]  
\includegraphics[scale=0.5]{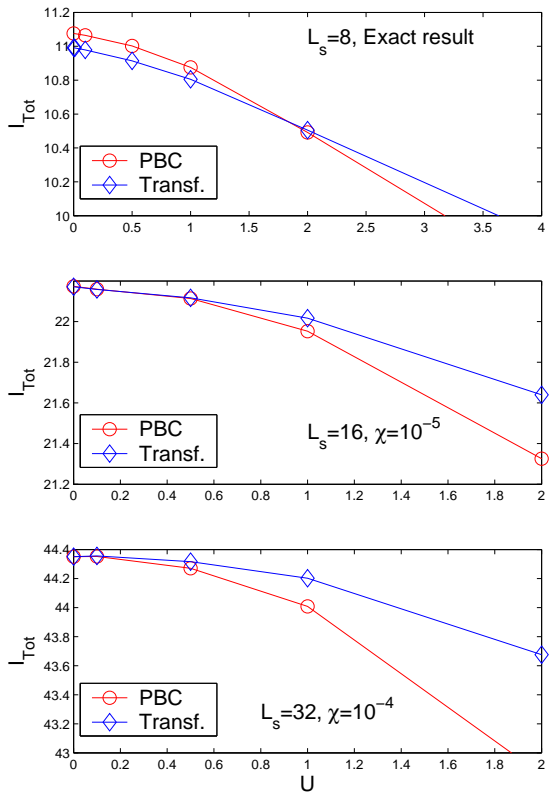}
\caption{Total quantum information for the original and 
transformed Hubbard models 
as a function of $U$ for various system sizes. 
For $L_{\rm_s}=8$ results are exact
while for larger system sizes data were obtained by setting 
$\chi=10^{-5}$.\label{fig:I-tot}}
\end{figure}

As seen from Fig.~\ref{fig:I-tot},
the total quantum information 
of the transformed Hubbard model~(\ref{eq:unitary})
in the geometry of Fig.~\ref{fig:hub-transf}~(a)
is smaller than that of the Hubbard model with periodic boundary conditions
only for very small values of the interaction strength,
$U< {\cal O}(t/L_{\rm s})$.
Apparently, the interactions of our transformed Hamiltonian, 
albeit fairly local, generate a strong entanglement between
lattice sites and subsystems because
they and the kinetic energy act as independent and actually
competing quantum channels. We shall investigate this point further
in section~\ref{sec:whywhywhy}. Here, we merely determine
the communication length by adding up equally
the distances for single-particle and two-particle electron transfers.
The terms $\hat{H}_{U, {\rm d}}$,  
$\hat{H}_{U, {\rm p}}$ and $\hat{H}_{U, {\rm s}}$ contribute equally
to give the estimate $C^{\rm transf}\approx 7 L_{\rm s}/2$. 
The larger entanglement in the transformed Hubbard model as expressed by
$C^{\rm transf}>C^{\rm pbc}=2(L_{\rm s}-1)$ 
implies that the DMRG allocates more computational resources
for the transformed Hamiltonian
than for the Hubbard model with periodic boundary conditions. 

{}From a technical point of view, the overall
CPU time increases by a factor of four to five also because
more matrix multiplications are necessary.
In contrast to the Hubbard model with periodic boundary conditions,
nine matrix multiplications instead of two 
must be carried out during the superblock diagonalization and 
three times more operators need to be renormalized. Moreover,
due to the new channels and the increase of entanglement in 
the system, the number of Davidson matrix multiplication increases
by a factor of two to three. 

\section{Effects of independent quantum channels}
\label{sec:whywhywhy}

In this section we study the entanglement generation in more detail.
To this end, we switch on 
perturbatively various coupling terms shown in Fig.~\ref{fig:1d-ham}.

\subsection{Smooth interpolation between open and periodic boundary conditions}

\begin{figure}[htb]
\includegraphics[scale=0.4]{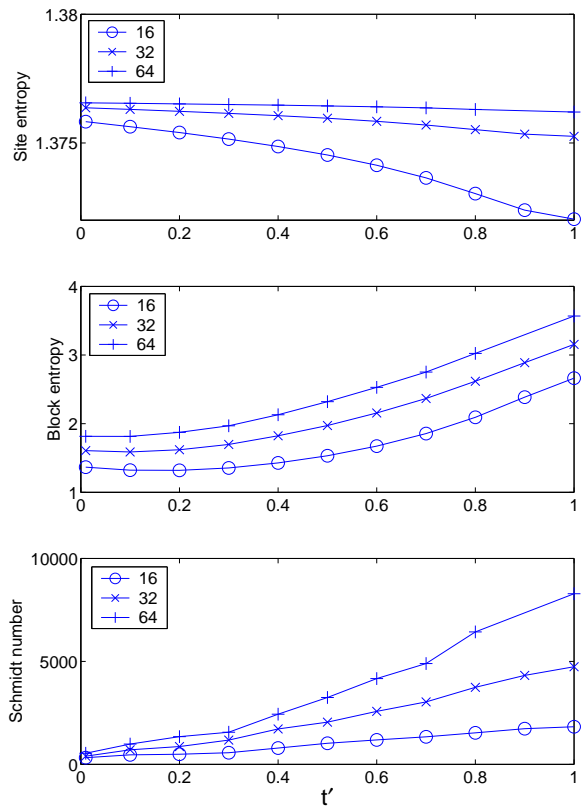}
\caption{Site entropy, block entropy,
and Schmidt number for the half-filled Hubbard model for the 
configuration shown in Fig.~\ref{fig:hub-order}~(a)
as a function of $t^{\prime}$ 
for $t=1$ and $U=0$ for $L_{\rm s}=16, 32$ sites, 
and $\chi=10^{-4}$, $M_{\rm min}=64$.\label{fig:hub-obc-pbc}}
\end{figure}

First, we consider how entanglement between DMRG blocks is generated  
for the configuration shown in Fig.~\ref{fig:hub-order}  
where we smoothly interpolate 
between open boundary conditions ($t^\prime=0$) and 
periodic boundary conditions ($t^\prime=1$) 
as a function of $t^{\prime}$ for $t=1$ and $U=0$. 
Our results are shown in Fig.~\ref{fig:hub-obc-pbc}. 
The total quantum information, 
the block entropy and the Schmidt number change smoothly as a function of
$t^{\prime}$. Thus, the second quantum channel opened by
the periodic boundary conditions behaves perturbatively. 
This is not always the case as seen in the next example.
Note, however, that the perturbation leads to a small effect only as long as
$t'<{\cal O}(t/L_{\rm s})$.

\subsection{Electron transfer between nearest and next-nearest neighbors}

\begin{figure}[tb]
\includegraphics[scale=0.33]{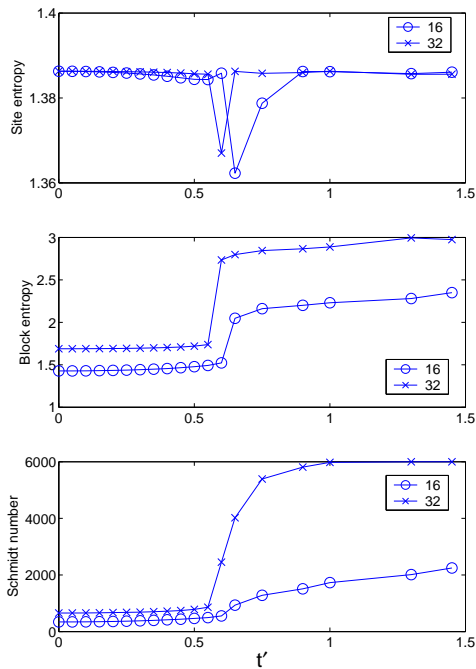}
\caption{Same as Fig.~\protect\ref{fig:hub-obc-pbc} 
for the half-filled Hubbard model
with open boundary condition as a function 
of the transfer integral~$t^{\prime}$ between next-nearest neighbors.
We set $t=1$ and $U=0$ for $L_{\rm s}=16, 32$ and $\chi=10^{-4}$, 
$M_{\rm min}=64$.\label{fig:hub-nnh}}
\end{figure}

Next, we analyze the entropy generation by an additional next-nearest 
neighbor electron transfer amplitude for the Hubbard model. 
The nearest-neighbor 
hopping is again set to $t=1$ whereas the transfer integral
between next-nearest neighbors $t^{\prime}$ is smoothly increased.
As in the previous example we choose $U=0$ to avoid 
\begin{figure}[hb]
\includegraphics[scale=0.33]{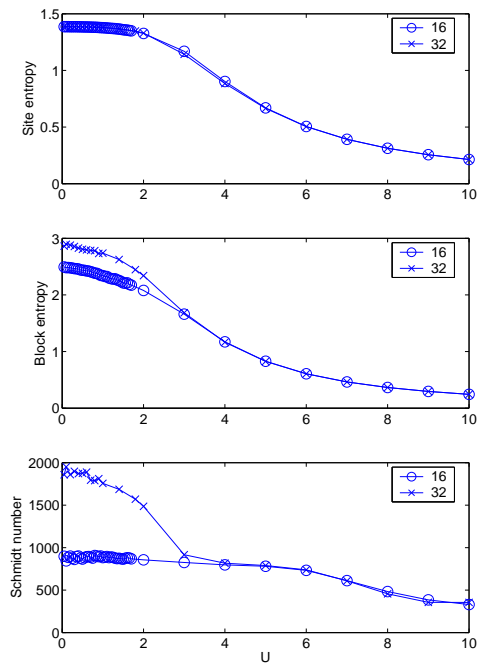}
\caption{Same as Fig.~\protect\ref{fig:hub-obc-pbc} 
but as a function of $U$ for the local-density term only,
$\hat{H}_{\rm d}=-\hat{H}_{\rm T}+U\hat{H}_{U, {\rm d}}$, 
as shown in Fig.~\protect\ref{fig:1d-ham}~(b).\label{fig:terms345}}
\end{figure}
the effect of other channels. In Fig.~\ref{fig:hub-nnh} 
we plot the site entropy, the block entropy
and the Schmidt number as a function of $t^{\prime}$.
These quantities
change smoothly as a function of $t^{\prime}$ up to a critical value 
where they increase rapidly. The value for the
rapid increase coincides
with the metal-insulator transition point 
in the $t$-$t'$-$U$ model, see, e.g., Ref.~[\onlinecite{noack-mi}].
For finite interaction strengths, the behavior of the entropies above
the quantum phase transition is more complex,
corresponding to the various phases of the $t$-$t'$-$U$ model~\cite{noack-mi}.

\subsection{Density-density interactions}
\label{subsec:densityinteractions}

We now turn to the effect of the interaction terms in the transformed
Hubbard model~(\ref{eq:unitary}).
We start with the analysis of the 
density-terms~$\hat{H}_{U, {\rm d}}$ in~(\ref{alltheUs}), as
shown in Fig.~\ref{fig:1d-ham}~(b). We neglect all 
other interaction terms and keep the single-particle hopping only,
as shown in Fig.~\ref{fig:1d-ham}~(a), i.e., we analyze
$\hat{H}_{\rm d}=-\hat{H}_{\rm T}+U \hat{H}_{U, {\rm d}}$.

As seen from Fig.~\ref{fig:terms345},
the site entropy, the block entropy and the Schmidt number 
do not change significantly as a function of $U$. Instead,
they mildly decrease as the interaction gradually eliminates 
double occupancies (and holes) from the Hilbert space.
For small interaction strengths, the block entropy and the Schmidt number
are fairly small. This is in accord with our observations for 
the Hubbard model with open boundary conditions.
Obviously, purely local density-type interactions 
do not open new quantum channels
and, therefore, they  do not substantial increase 
\begin{figure}[hb]
\includegraphics[scale=0.33]{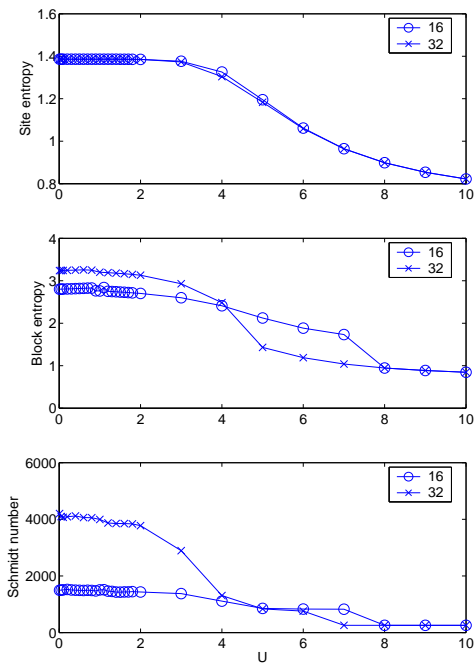}
\caption{Same as Fig.~\protect\ref{fig:hub-obc-pbc} 
but as a function of $U$ for the pair-hopping term only,
$\hat{H}_{\rm p}=-\hat{H}_{\rm T} +U\hat{H}_{U, {\rm p}}$, 
as shown in Fig.~\protect\ref{fig:1d-ham}~(c).\label{fig:terms67}}
\end{figure}
the entanglement between blocks.
Density-density interactions between different lattice sites 
behave qualitatively the same because they do not involve the exchange
of particles between the blocks. Therefore, the DMRG still performs well
for the Hubbard model with long-range density-density interactions 
when open boundary conditions are applied.

The situation changes when we treat the pair-hopping 
term~$\hat{H}_{U, {\rm p}}$~(\ref{alltheUs}), 
shown in Fig.~\ref{fig:1d-ham}~(c),
together with the kinetic energy $\hat{H}_{\rm T}$. We ignore
all other interaction terms, i.e., we treat
$\hat{H}_{\rm p}=-\hat{H}_{\rm T}+U \hat{H}_{U, {\rm p}}$ 
in Fig.~\ref{fig:terms67}.
Again, the site entropy, the block entropy and the Schmidt number
decrease smoothly as a function of the interaction strength.
In comparison with the purely local interaction~$\hat{H}_{U, {\rm d}}$
the Schmidt number has almost doubled.

\begin{figure}[tb]
\includegraphics[scale=0.33]{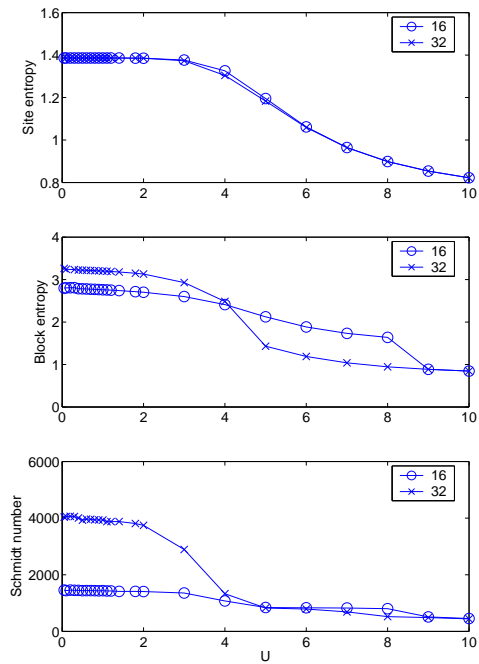}
\caption{Same as Fig.~\protect\ref{fig:hub-obc-pbc} 
but as a function of $U$ for the pair-hopping term only,
$\hat{H}_{\rm s}=-\hat{H}_{\rm T} +U\hat{H}_{U, {\rm s}}$, 
as shown in Fig.~\protect\ref{fig:1d-ham}~(d).\label{fig:terms89}}
\end{figure}

The increase in block entropy and Schmidt number is 
very similar when we study the effect of the spin-flip 
term~$\hat{H}_{U, {\rm s}}$ in~(\ref{alltheUs}), 
as shown in Fig.~\ref{fig:1d-ham}~(d). The result of the analysis
of $\hat{H}_{\rm s}=-\hat{H}_{\rm T}+U \hat{H}_{U, {\rm s}}$
is shown in Fig.~\ref{fig:terms89}.
Apparently, the spin exchange between 
neighboring sites creates entanglement similar to
the exchange of pairs. In comparison of the effect of 
$\hat{H}_{U, {\rm d}}$
on the one-hand side and
$\hat{H}_{U, {\rm p}}$, $\hat{H}_{U, {\rm s}}$ on the other we
conclude that not only the 
number of links and their strength but also the type of
coupling plays an important role for the entanglement.

\subsection{Super-site representation}
\label{subsec:supersite}

\begin{figure}[tb]
\includegraphics[scale=0.33]{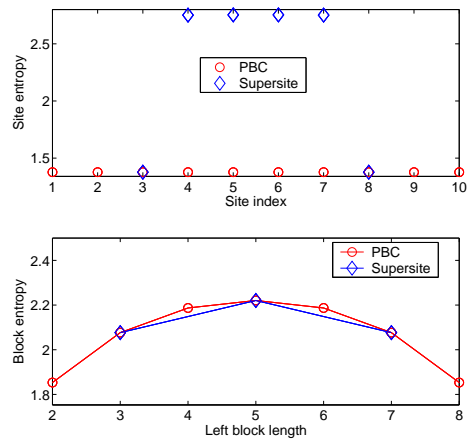}
\caption{Site entropy and block entropy  
for the half-filled Hubbard model for periodic boundary conditions
shown in Fig.~\protect\ref{fig:hub-order}~(a) and
for the super-site representation in Fig.~\protect\ref{fig:hub-transf}
for $U=1$ and $\chi=0$ (exact calculation).\label{fig:super-site}}
\end{figure}

Originally, as shown in Fig.~\ref{fig:hub-transf},
the transformed Hubbard model~(\ref{eq:unitary})
is defined on a two-chain geometry with purely local interaction
and electron transfers between neighboring sites. 
One may wonder whether the analysis of 
subsection~\ref{subsec:densityinteractions} is adequate
because it is based on the single-chain geometry of Fig.~\ref{fig:1d-ham}.

In order to clarify this issue, we
reduce the number of quantum channels between the DMRG blocks
by forming `super-sites' from lattice sites of type~$a$ and $b$. 
In this representation a lattice with $L_{\rm s}/2-1$ sites and
$q=16$ degrees of freedom per site is formed, plus two end sites
with $q=4$ degrees of freedom, and open boundary conditions apply.

\begin{figure}[htb]
\includegraphics[scale=0.33]{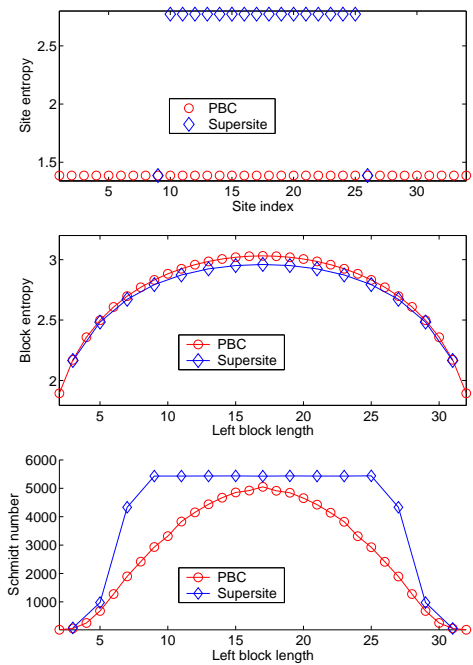}
\caption{Same as Fig.~\ref{fig:super-site}
for $U=0$ for $N=34$, $\chi=10^{-4}$.\label{fig:super-site-n34}} 
\end{figure}

The two sites at the boundaries have the same site entropy for
both models. In the super-site representation we halve the length of the system
so that sites in the interior of the chain carry an entropy which is
twice as large as for sites in the Hubbard model with periodic 
boundary conditions.    
A comparison of the block entropies is only meaningful for blocks which
contain the same number of sites, i.e., $l=3,5,7$ in Fig.~\ref{fig:super-site}.
As expected, for these block lengths they agree for both models.
The total quantum information is essentially the same for both configurations.
This observation is again readily explained by the fact that there are 
two quantum channels between the DMRG blocks in both 
representations. 
This is shown explicitly 
in Fig.~\ref{fig:super-site-n34} for $U=0$
and $N=34$, $\chi=10^{-4}$ using maximum $M=350$ block states.

The advantage that the transformed 
Hubbard model in the super-site representation
is only of length $L_{\rm s}/2+1$ is more than compensated by the fact 
that during the numerical calculation there are $q=16$ degrees of freedom
for the two intermediate sites $s_l$ and $s_r$ 
in the superblock representation.
This higher demand for computational resources can be reduced if instead of two
intermediate sites in the superblock representation only a single site is used.
The modification of the DMRG 
in this direction is possible~\cite{Verstrate-pbc,white-one-site}. 

\section{Conclusions}
\label{sec:conclusions}

We have constructed a unitary transformation between basis states 
for the one-dimensional Hubbard model subject to periodic boundary conditions
so that open boundary conditions apply for the transformed Hamiltonian. 
Despite the fact
that the one-particle and two-particle interaction matrices link nearest
and next-nearest neighbors only, the performance of the density-matrix
renormalization group method for the transformed Hamiltonian
does not improve significantly because some of the new interactions
act as independent quantum channels which generate the same level
of entanglement as periodic boundary conditions in the original formulation
of the Hubbard model. 

The total quantum correlation in the system 
for the transformed model decreases
only for small interaction strengths 
and for very short chain lengths. Therefore, this
approach cannot be used to improve the performance of the DMRG for
reasonable system sizes. 
We have shown that the localization of interactions alone does not improve
the performance of the DMRG. Instead, it is affected more significantly 
by the number and the strength of the various quantum 
channels between the DMRG blocks.  

In conclusion, our results contribute to a better understanding 
of the entanglement production within the DMRG. We propose to implement the
communication length for the construction of an optimal
basis and the proper ordering of lattice sites. The expected
reduction of the block entanglement should improve 
the performance of the DMRG for Hamiltonians 
with long-range electron transfers and interactions.

\mbox{}
\vspace*{-24pt}
\acknowledgments

This research was supported in part by the Hungarian Research Fund (OTKA)
Grants No.~T~043330 and F~046356. The authors acknowledge
computational support from Dynaflex Ltd.\ under Grant No.~IgB-32.
\"Ors Legeza was also supported by a J\'anos Bolyai Research Scholarship.


\begin{thebibliography}{99}

\bibitem{White}
S.R.\ White, Phys.~Rev.~Lett.~{\bf 69}, 2863 (1992); 
Phys.\ Rev.\ B~{\bf 48}, 10345 (1993).

\bibitem{Xiang} T.~Xiang, Phys.\ Rev.\ B~{\bf 53}, 10445 (1996).  

\bibitem{Gebhard-kdmrg} S.~Nishimoto, E.~Jeckelmann, F.~Gebhard, 
and R.M.\ Noack, Phys.\ Rev.\ B~{\bf 65}, 165114 (2002).

\bibitem{Legeza-ms} \"O.~Legeza and J.~S\'olyom, 
Phys.\ Rev.\ B~{\bf 68}, 195116 (2003).

\bibitem{White-qc} S.R.~White and R.L.\ Martin, 
J.~Chem.\ Phys.~{\bf 110}, 4127 (1998); 
S.~Daul, I.~Ciofini, C.~Daul, and S.R.\ White, 
Int.\ J.\ Quantum Chem.~{\bf 79}, 331 (2000).

\bibitem{Chan-qc} G.K.-L.\ Chan and M.\ Head-Gordon, 
J.\ Chem.\ Phys.~{\bf 116}, 4462 (2002).

\bibitem{Mitrushenkov-qc} A.O.\ Mitrushenkov, G.\ Fano, F.\ Ortolani, 
R.\ Linguerri, and P.\ Palmieri, 
J.\ Chem.\ Phys.~{\bf 115}, 6815 (2001).

\bibitem{Legeza-dbss} \"O.\ Legeza, J.\ R\"oder, and B.A.\ Hess, 
Phys.\ Rev.\ B~{\bf 67}, 125114 (2003).

\bibitem{Chan-qc1} G.K.-L.\ Chan and M.\ Head-Gordon, 
J.\ Chem.\ Phys.~{\bf 118}, 8551 (2003).

\bibitem{Legeza-qc} \"O.\ Legeza, J.\ R\"oder, and B.A.\ Hess, 
Mol.\ Phys.~{\bf 101}, 2019 (2003).

\bibitem{Noack1} R.M.\ Noack and S.R.\ White in 
{\sl Density Matrix Renormalization: A New Numerical Method in Physics}, 
ed.\ by I.\ Peschel, X.\ Wang, M.\ Kaulke, and K.\ Hallberg
(Springer, Berlin, 1999), p.~27.

\bibitem{Dukelsky} J.~Dukelsky and S.~Pittel, 
Rep.\ Prog.\ Phys.~{\bf 67}, 513 (2004).

\bibitem{Schollwoeck} U.~Schollw\"ock, Rev.\ Mod.\ Phys.~{\bf 77}, 259 (2005).

\bibitem{Noack2} R.M.~Noack and S.R.~Manmana in {\sl Lectures on the phy\-sics 
of highly correlated electron systems~IX}, ed.~by A.\ Avella and
F.~Mancini (AIP Conference proceedings~{\bf 789}, Melville, New York, 2005),
p.~93.

\bibitem{Cuthill} E.\ Cuthill and J.\ McKee in 
{\sl Proceedings of the 24th National Conference of the 
Association for Computing Machinery} (ACM Press, New York, 1969), p.~157.

\bibitem{Legeza-qdc} \"O.~Legeza and J.~S\'olyom, 
Phys.\ Rev.\ B~{\bf 70}, 205118 (2004).

\bibitem{Vidal} J.\ Vidal, G.\ Palacios, and R.\ Mosseri, 
Phys.\ Rev.\ A~{\bf 69}, 022107 (2004).

\bibitem{Korepin} H.\ Fan, V.\ Korepin, and V.\ Roychowdhury, 
Phys.\ Rev.\ Lett.~{\bf 93}, 227203 (2004).

\bibitem{Rissler1} J.\ Rissler, R.M.\ Noack, and S.R.\ White, unpublished
(cond-mat/0508524, 2005).

\bibitem{Moritz-qc} G.~Moritz, B.A.~Hess, M.\ Reiher,
J.\ Chem.\ Phys.~{\bf 122}, 024107 (2005).

\bibitem{Verstrate-pbc} F.\ Verstraete, D.\ Porras, and J.I.\ Cirac,
Phys.\ Rev.\ Lett.~{\bf 93}, 227205 (2004).

\bibitem{Legeza-deas} \"O.\ Legeza and J.\ S\'olyom, (unpublished, 2005).

\bibitem{Bethe} For a review, see F.H.L.~Essler,
H.~Frahm, F.~G\"ohmann, A.~Kl\"umper, and V.E.~Korepin,
{\sl The one-dimensional Hubbard model} (Cambridge University Press, 
Cambridge, 2005).



\bibitem{noack-mi} G.I.~Japaridze, R.M.~Noack, and D.~Baeriswyl 
(unpublished, 2005). 

\bibitem{white-one-site} 
S.R.\ White, unpublished (cond-mat/0508709, 2005).
\end{thebibliography}
\end{document}